\documentclass[10pt,twocolumn]{IEEEtran}
\usepackage{mathptmx}
\usepackage{multicol}

\usepackage{color}
\usepackage{makecell}
\usepackage{xcolor}
\usepackage{soul}
\usepackage{amsmath}
\usepackage{graphics}
\usepackage{setspace}
\usepackage{cite}
\usepackage{latexsym}
\usepackage{float}
\usepackage{epsfig}
\usepackage{multirow}
\usepackage{cite,cases,url}
\usepackage{amssymb}
\usepackage{graphicx}
\usepackage{epstopdf}
\usepackage{fancyhdr}
\usepackage{afterpage}
\usepackage{balance}
\usepackage{subcaption}
\usepackage{enumerate}
\usepackage{array}
\usepackage{algorithmic}
\usepackage{algorithm}
\usepackage{enumitem}
\usepackage{dblfloatfix}
\usepackage{tabularx}
\usepackage{siunitx}
\usepackage[normalem]{ulem}
\useunder{\uline}{\ul}{}
\newcolumntype{Y}{>{\centering\arraybackslash}X}

\newcolumntype{L}{>{\centering\arraybackslash}m{5cm}}
\newcolumntype{K}{>{\centering\arraybackslash}m{6cm}}
\newcolumntype{P}{>{\centering\arraybackslash}m{2.3cm}}
\newcolumntype{M}{>{\raggedright\arraybackslash}m{2cm}}
\newcolumntype{N}{>{\raggedright\arraybackslash}m{2.5cm}}

\fancypagestyle{firstpage}
{
    \fancyhead[L]{\footnotesize This article has been accepted for publication in the IEEE Communications Magazine. Please cite it as A. S. Abdalla, A. Yingst, K. Powell, A. Gelonch-Bosch, and V. Marojevic, ''Open-Source Software Radio Platform for Research on Cellular Networked UAVs -- It Works!,'' IEEE Communications Magazine, March 2022. } 
    \fancyhead[R]{\footnotesize \thepage}
    \cfoot{}
    
}

\begin{document}


\title{{Open-Source Software Radio 
Platform for 
Research on Cellular Networked 
UAVs -- It Works!
}}

\author{
\IEEEauthorblockN{Aly Sabri Abdalla, Andrew Yingst, Keith Powell,
Antoni Gelonch-Bosch, and Vuk Marojevic
}
}

\maketitle
\thispagestyle{firstpage}
\begin{abstract}
Cellular network-connected unmanned aerial vehicles (UAVs) 
experience different radio propagation conditions than radio nodes on the ground. Therefore, it has become critical to investigate the performance of aerial radios, both theoretically and through field trials. In this paper, we
consider low-altitude aerial nodes that are served by an experimental 
cellular network. We provide a detailed description of the 
hardware and software components needed for establishing a broadband wireless testbed for UAV communications research using software 
radios. 
Results show that a testbed for innovation in UAV communications and networking is feasible with commercial off-the-shelf hardware, open-source software, and low-power 
signaling. 
\end{abstract}
\IEEEpeerreviewmaketitle

\vspace{-2mm}
\section{Introduction}
\label{sec:intro}
\begin{figure*}[t]
\begin{center}
\includegraphics[width=0.75\textwidth]{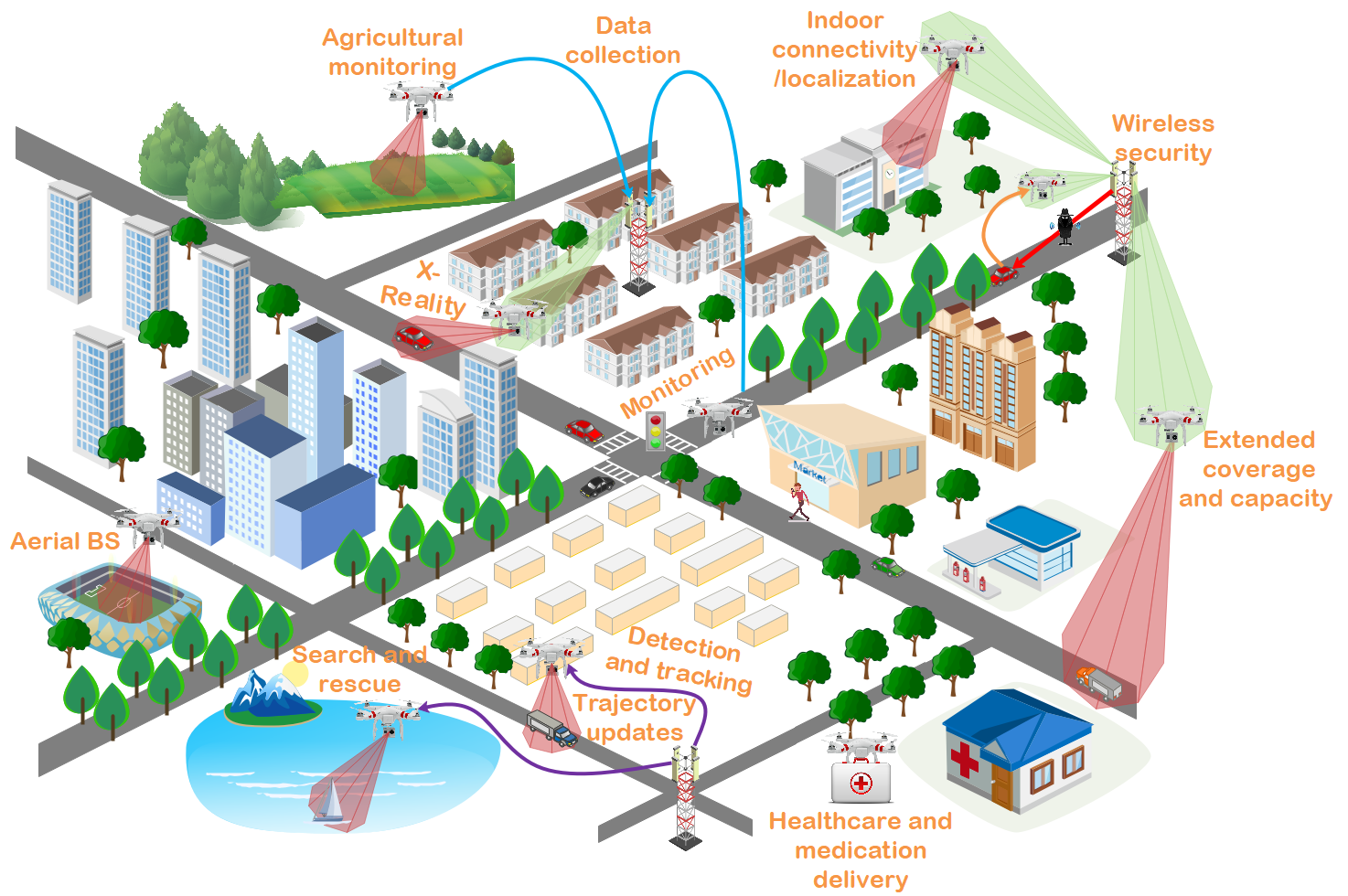}
\end{center}
\caption{Emerging applications and use cases for networked UAVs. 
}
\vspace{-5mm}
\label{fig:Figurex}
\end{figure*}

Unmanned aerial vehicles (UAVs) are increasingly popular in the commercial sector and are considered one of the industrial verticals of the fifth generation of mobile communications (5G). 
UAVs can support intelligent transportation systems through traffic monitoring, accident reporting, and 
aerial delivery of 
cargo and medication, among others. 
Among many applications, UAVs can 
monitor crops, detect weeds, gather 
sensor data, and assist with search and rescue 
and 
disaster recovery operations 
\cite{R4_SAR}. 
Fig.~\ref{fig:Figurex} illustrates this. 

The characteristics 
that make a UAV particularly suitable for the above-mentioned applications are the strong line of sight (LOS) 
links, 
 3D mobility, 
low cost, and ability to operate in a hostile environment. 
The cellular network performance 
can be improved with UAVs by 
implementing stronger security mechanisms~\cite{AlySecurity}, extending the coverage, and improving the capacity 
~\cite{R4_connectivity}. This is being standardized by The Third Generation Partnership Project (3GPP)~\cite{3GPP_Stnds}. 

The 3GPP Technical Report (TR)~36.777 and Technical Specification (TS)~22.125 define 
the essential requirements for providing unmanned aircraft system (UAS) services through long-term evolution (LTE) networks. TR~23.754 and TR~23.755 identify the network infrastructure and the application architecture to support UAV-enabled applications and use cases in 5G networks. \textcolor{black}{TR~33.854 specifies the requirements and procedures to provide secure communications services to UASs served by 3GPP networks.} The International Telecommunication Union (ITU) 
and the IEEE 
provide complementary standards for UAV applications and use cases \cite{R3,IEEE1920}.

UAVs are not only considered as user equipment (UEs) in the cellular network, but also as network support nodes.
They have the potential to facilitate reliable, secure, and efficient wireless access~\cite{TVT}. A large number of applications involving UAVs and advanced wireless communications are expected to be developed in the coming years. 
To realize this vision, it is necessary to have flexible research platforms and testbeds that are widely accessible. \textcolor{black}{This will accelerate 
research and development (R\&D) that will provide practical insights on the specific services that cellular networks should offer to enable safe and efficient operation and cooperation among UASs. The 3GPP provides initial performance requirements for the network to support command and control and video transmission from UAVs~\cite{3GPP_Stnds}. Cellular network technology and procedures need to evolve to, among others, improve the 3D coverage and effectively handle the challenging interference scenario between UAV and terrestrial UEs.}

Software defined radios (SDRs) can be used 
with a UAV for rapid radio prototyping and experimentation. 
\textcolor{black}{An SDR does the radio frequency (RF) data acquisition/generation and data conversion 
for waveforms and protocols to be processed 
in software \cite{alex16sdr}}.
\textcolor{black}{Using a pair of Universal Software Radio Peripherals (USRPs) and open-source LTE software srsRAN, \cite{srsLTEthroughput} 
reports average throughput and signal-to-noise ratio values at 10, 25, and 50 m UAV altitudes from the ground SDR base station (BS) as 5.5, 2.0, 4.5 Mbps and 10.0, 7.5, 9.0 dB, respectively, for a 10 MHz frequency-division duplex (FDD) LTE system at 3.7 GHz.}
\textcolor{black}{Those values are rather low, even without an external power amplifier (PA). This may stem from choosing too high USRPs gains resulting in signal distortion and self-interference from 
the design lacking isolation between the transmit and receive 
antennas
~\cite{srsLTEthroughput}}. 
Another 
trial evaluates the performance of the UAV as a relay between the ground BS and UE~\cite{Rebots}. 
The open-source OpenAirInterface (OAI) software with a USRP on board the UAV provides the experimental LTE 
aerial relay (AR). 
The measurements show that the 
throughput slightly improves 
when the signals are relayed through the 
AR, 
specifically for 
non-LOS 
links between the BS and the UE.
\textcolor{black}{In another research project~\cite{SkyCell}, the UAV with a USRP and srsRAN is deployed as an aerial BS (ABS) that positions itself to 
maximize network throughput for three ground UEs. 
Throughput measurements were conducted in an outdoor drone cage 
over short distances.}   

\textcolor{black}{
The published SDR platforms' evaluation results} often do not provide the necessary performance because of poor design or operating decisions. The difficulty is to select, integrate, and configure the hardware and the 
software for providing a flexible and accessible platform that can provide near commercial-grade 
performance and that can be scaled for enabling cellular-connected UAV research in a production-like environment.

\textcolor{black}{The purpose of this paper is to identify the requirements, the design tradeoffs, the components and practical design and configuration choices for building a reproducible open-source SDR platform that provides the necessary interfaces and performance figures 
for enabling unlimited research opportunities with cellular network-connected UAVs. The proposed platform is the basis for the first 
SDR experiments offered through the \textit{Aerial Experimentation and Research Platform for Advanced Wireless} (AERPAW) \cite{aerpaw19}.} 


The rest of the paper is organized as follows: 
Section II introduces the platform requirements. 
Section III 
assesses the design options, tradeoffs, and 
component choices. We show early results  and offer guidelines for research in this area. 
Section IV discusses the key challenges and directions for advanced wireless R\&D with UAVs. 
Section V provides the concluding remarks. 

\section{\textcolor{black}{Requirements of 
cellular network-connected UAV research platforms}} 
The high demand for UAV integration into cellular networks 
requires a research platform that can evolve with technology. Deploying a production-like 4G/5G experimental cellular network requires implementing the radio access network (RAN), the backhaul, and the core network (CN). This is the goal of {AERPAW}, 
a public-private partnership project that is developing and operating a large-scale testbed for advanced wireless research with UAVs.
The requirements for the open-source SDR research platform 
are

\begin{itemize}[leftmargin=+9.4pt]
    \item 3GPP standard-compliant cellular communications system that performs according to industry benchmarks,
    \item Flexible and modular radio interface, supporting various cellular network-connected UAV experiments, 
    \item Effective integration of the radio system with the UAV, subject to the UAV's space, weight and power constraints, 
    \item Open-source software that is maintained and that implements the RAN, the backhaul, the CN, and the UEs,
    \item Commercial off-the-shelf (COTS) hardware that is widely available and that is popular among researchers,
    \item Inter-operable and well-supported software and hardware, 
    \item Reproducible platform and experiments,
    \item Open application programming interfaces (APIs),
    \item Full control over the experiment (radio and vehicle), and
    \item Scalable and portable platform that facilitates integration with 
    other testbeds, vehicles, or infrastructure.
\end{itemize}
\vspace{2mm}


The above requirements translate into specific requirements for the different building blocks of the platform. These are summarized in Table I and elaborated in continuation. 

\noindent\textbf{Data Conversion:} 
The wireless transmission and reception of 
software-defined waveforms requires data sampling and up/down conversion. These processes are facilitated by commercial SDR hardware, which 
should support commercial 4G and 5G sampling rates, instantaneous bandwidths of at least 20 MHz, and frequency agility. 
Moreover, it should allow multi-channel communications to support transmit (Tx) and receive (Rx) diversity, multiple-input multiple-output (MIMO) communications, and beamforming. 

\noindent\textbf{RF Front End:}
The goal is to have a modular and flexible RF front end for supporting multiple bands, providing enough RF transmission power and filtering to comply with spectrum regulations, Tx-Rx antenna isolation to avoid receiver saturation, 
and enough gain to recover weak signals at the limit of the coverage area, initially defined as 1 km from the BS. 
The choice of RF components will eventually be driven by the 
experimenter needs and the commercial availability of RF components 
for the desired frequency. 
Obtaining a local license to radiate from the ground and air is also necessary. 


\noindent{\textbf{Processing Unit:}}
The baseband processing has the highest computing demand of a 
cellular communications system. It can be executed on a general-purpose processor with sufficient processing power.  
The processor also needs to provide 
several network interfaces for data and control. 



\noindent\textbf{Power:} Computers, SDRs, and active RF components need a power supply. 
The power requirements are defined by the RF and processing units, 
experiment configuration, 
and 
experiment duration. 

\noindent\textbf{SDR Software:}
{AERPAW} proposes to leverage popular open-source software libraries that implement modern RANs, CNs, and UEs. The reason is that the testbed is built for enabling global and cutting-edge research, which can be excelled when the code is open, maintained, and easy to modify. 
This enables rapid prototyping and experimenting variations to the RF signaling and protocol processing.
The main criteria for choosing a software library are its user community, COTS hardware compatibility and portability, and interoperability with other experimental or production radio equipment. 

\noindent\textbf{Experimental and Control Networks:}
The experimental network connects UEs to the RAN via SDRs and the RAN to the CN via radio, fiber, or copper. 
All BSs need IP connectivity to the 
CN to implement the S1 interface for authentication, mobility management, and so forth. 
AERPAW users submit their experiments and, when approved, the experiment runs in batch mode. 
However, for initial platform development and during regular operation, experiments are controlled by the testbed operator, who needs to be able to configure, start, stop, or pause an experiment as needed, e.g. when undesired situations are detected. 
Being able to connect to each SDR computer via secure shell (ssh) is the goal for initial development, testing, and experiment control. 

\noindent\textbf{UAS:}
The UAS is the UAV and its remote controller (RC). Unless a UAV flight waiver is obtained, a certified pilot must be in visual LOS with the UAV. 
The UAV needs to be robust enough to lift the experimental radio system payload.  
An experiment may encompass evaluating or choosing a UAV  trajectory 
as a function of the observed radio parameters. Timestamped and location-specific data, typically provided by the onboard global positioning system receiver, therefore need to be logged and provided to the experimenter after 
the experiment.
\begin{table*}
\centering
\caption{Requirements and proposed design for the SDR platform enabling \textcolor{black}{4G/5G} cellular networked UAV research.}
\footnotesize
\label{tab:Table_combined}
\centering
{\begin{tabular}{|p{0.6cm}|p{1.2cm}|p{6.8cm}|p{8cm}|}
\hline
\multicolumn{2}{|c|}{\textbf{System Features}}  &\textbf{Requirements } &  \textbf{Components} 
\\
\hline
{
SDR}
& 
Data conversion 
&
\textcolor{black}{- {COTS SDR hardware with drivers that support sampling rates of modern wireless systems, permit a wide range of RF, and have broad research community support.}} 
& 
- B205mini-i USRP: Up to 61.44 MHz sampling rate, 56 MHz instantaneous bandwidth, 
0 dBm maximum Rx power, low Tx output power \cite{Keith_ACM20}, $<$8 dB noise figure (NF), 70 MHz to 6 GHz. 
\\
\cline{2-4}

& 
{Baseband/ protocol processing} 
& 
\textcolor{black}{- {Multi-core x86 computer with 3 GHz or higher clock per core (at least 4 physical cores) 
\textcolor{black}{with USB 3.0 or 10 Gbps NIC}. 
}} 
&
- {Dell OptiPlex 7070 Micro} (fixed node): Intel i9-9900 8-core processor, 32 GB RAM, \textcolor{black}{USB3, two-port 10 Gbps NIC (SFP+), 1 Gbps NIC.} 
\newline - {Intel \textcolor{black}{NUC10 (NUC10i7FNK)}} (mobile node): Intel i7-10710U 4-core processor, 32 GB RAM, \textcolor{black}{USB3, Thunderbolt 3, 1 Gbps NIC}. 
\\
\cline{2-4}

& 
Software  
& 
\textcolor{black}{- {Supported open-source software libraries;
closed source software can be useful, e.g. for baseline experiments.}} 
& 
- The latest USRP Hardware Driver (UHD) 4.0, 
srsRAN version 20.10.1, open5GS version 2.2.9, and  
OAI version 1.2.2, are 
installed on Ubuntu 18.04 
in Docker containers.
\\
\hline
{
RF Front End} 
& 
\rule{0pt}{3ex} Amplifier 
& 
\textcolor{black}{- Transmit PA 
for desired 1 km range; gain and other characteristics depend on the SDR and the output power it can cleanly produce for a wideband signal.}

\textcolor{black}{- 
Low power consumption, weight, size, and cooling for mobile node PA.}

\textcolor{black}{- Receiver LNA 
to improve reception range.}

&
- {Mini-Circuits (MC) ZHL-15W-422-S+}
PA: 15 W, 46 dB gain, 38-40~dBm output power at 1 dB compression, 49 dBm OIP3, 10 dB NF, 0.6-4.2 GHz. 
\newline - {MC ZVE-8G+} (mobile Tx) PA: 1 W, 34-36 dB gain, 32 dBm output power 
at 1 dB compression, 40 dBm OIP3, 4-5 dB NF, 2-8 GHz. 
\newline - {MC ZX60-83LN12+} LNA (Rx): 22 dB gain, 13-20 dBm output power at 1 dB compression, 30-35 dBm OIP3, 1.4-2.3 NF, 0.5-8 GHz.
\\

\cline{2-4}
& \rule{0pt}{4ex}    {Filters}  & 
\textcolor{black}{- Eliminate harmonics and spurs in the transmitter.} 

\textcolor{black}{- Limit the amount of undesired signals entering the receiver, yet provide frequency agility.} 
&
- {MC VLF-4400+} (Tx): low pass filter DC-4400 MHz passband (PB), 3 dB loss at 5290 MHz. 
\newline - MC VBFZ-3590+ (Rx): 
bandpass filter with 3000-4300 MHz PB. 
\\
\cline{2-4}
&
Antennas
&
\textcolor{black}{- Broad coverage and frequency agility.}

\textcolor{black}{- Small and lightweight for the UAV node.}

&
- {Mobile Mark RM-WB1-DN-B1K} (fixed node): 617-960 and 1700-6000 MHz, 3 dBi peak gain, up to 10 W.
\newline - {Octane Wireless SA-1400-5900} (mobile node): 1400-5900 MHz, 
up to 500 mW.
\\
\hline
{
Net-work} & 
Experimen-tal network  
& \textcolor{black}{- {Reliable interface between the RAN and the CN.} 
}
&
- 
\textcolor{black}{Ethernet or} WiFi6 (AX4800 router) for the backhaul 
and BS-BS links \textcolor{black}{(X2 interface)}. 
\\
\cline{2-4}

& 
Experiment control  & 
\textcolor{black}{- {
Reliable wireless side channel for aerial SDR experiment control.} 
} 
&
- Private IP network 
using 
ssh via WiFi6.
\\
\hline
{
UAS} &
{UAV}& 
\textcolor{black}{- {
Easy takeoff and landing, hovering capability, and 5~kg payload capacity.} 
} &
- DJI Matrice 600: Hexacopter with 5.5 kg maximum payload and 6~x~5,700 mAh  battery capacity (TB48S battery pack). 
\\
\cline{2-4}
&
{UAV control}& 
\textcolor{black}{- {Certified UAV pilot with RC and visual line of sight to the UAV due to current regulations.} 
} &
- DJI Matrice 600 manufacturer supplied RC. On-site operator, one per UAV.
\\
\hline
{
Power} & Ground node    &
{
\textcolor{black}{- 
Direct AC, if available, or a generator for powering computers, SDRs and RF components. }}& 
- 2000 W portable gasoline generator 
with grounding rod collocated with each fixed node at the 
test location.
\\
\cline{2-4}

& Aerial node &
{\rule{0pt}{2ex}
\textcolor{black}{- Autonomous (battery-powered) operation.}} 


&
- Separate power sources for UAV and SDR payload. 

- The UAV comes with high-capacity batteries (30 minute flight).

- 5S 3000 mAh LiPo battery 
for NUC10 + SDR, PA, and LNA. 
\\
\hline
\end{tabular}
\vspace{-3 mm}
}
\end{table*}

\section{Software and hardware options and the system design tradeoffs to make it all work} 
\textcolor{black}{The selection of components depends on their compatibility, ease of integration, flexibility, scalability, performance, and availability. 
While we organize this section similar to the previous, the software (SW) and hardware (HW) interactions are critical 
for the choice of components to prototype an efficient platform that meets the outlined requirements. Table I lists the specific components \textcolor{black}{of the proposed platform}.}

\begin{figure}[t]
    \centering
    \includegraphics[width=0.49\textwidth]{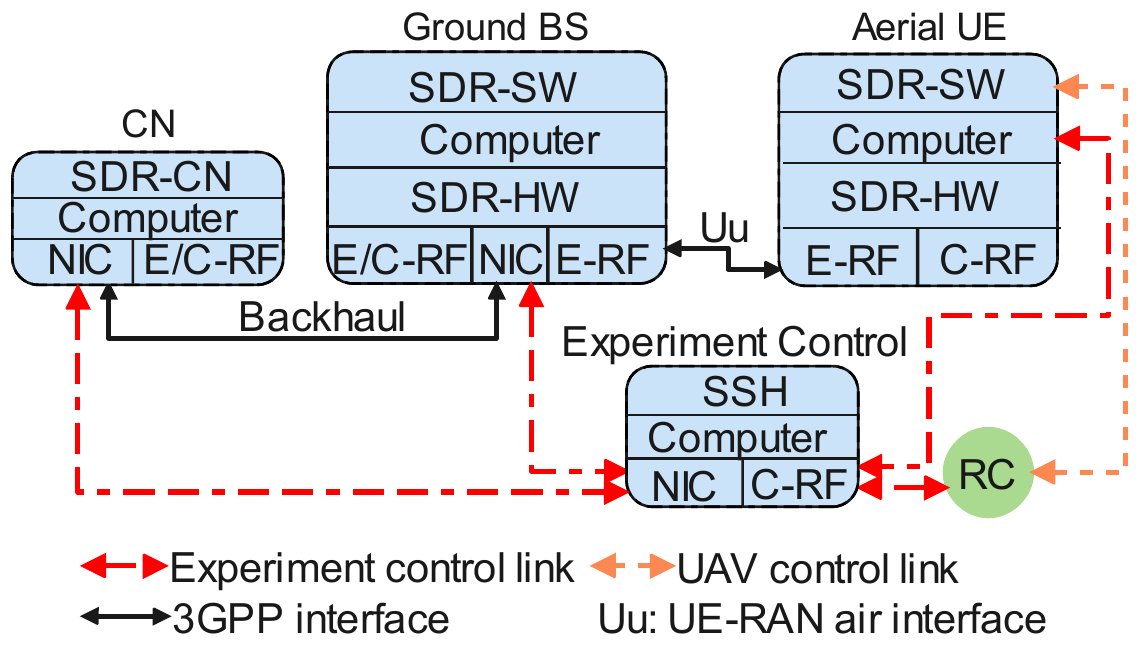}
    \caption{\textcolor{black}{
    Testbed interfaces (E/C-RF -- experiment/control RF; NIC -- network interface card
    ).
    }}
    
    \label{fig:Figure3}
    \vspace{-3mm}
\end{figure}

\textcolor{black}{Fig.~\ref{fig:Figure3} shows the minimum necessary system components and the logical interfaces. Defining the components and interfaces facilitate scalability 
and portability. 
} 


\subsection{Radio Network Hardware}


\noindent\textbf{Data Conversion:} 
\textcolor{black}{Among the examined options, USRPs offer several advantages: a wide set of board and RF module options, 56+ MHz instantaneous bandwidth per channel, and a single open-source hardware driver. 
There are lightweight and small form factor devices that fit well on a UAV. 
We chose USRPs mainly because of their flexibility and openness, their popularity among the broader R\&D community, and 
their compatibility with advanced SDR software, the popular GNU Radio framework, and Matlab. 
USRPs have their limitations, including low Tx power, especially for wideband multicarrier waveforms, limited analog filtering, and interoperability issues among certain driver versions, devices, software, and frequencies. These can be overcome 
by careful hardware and software design, configuration, and systematic laboratory and outdoor testing. We use auxiliary SDR hardware, RF instruments, and open-source tools, such as iPerf, ping, Wireshark, and SDR-specific tools for signal and network analysis.}

\begin{figure}[t]
\begin{center}
\includegraphics[width=0.49\textwidth]{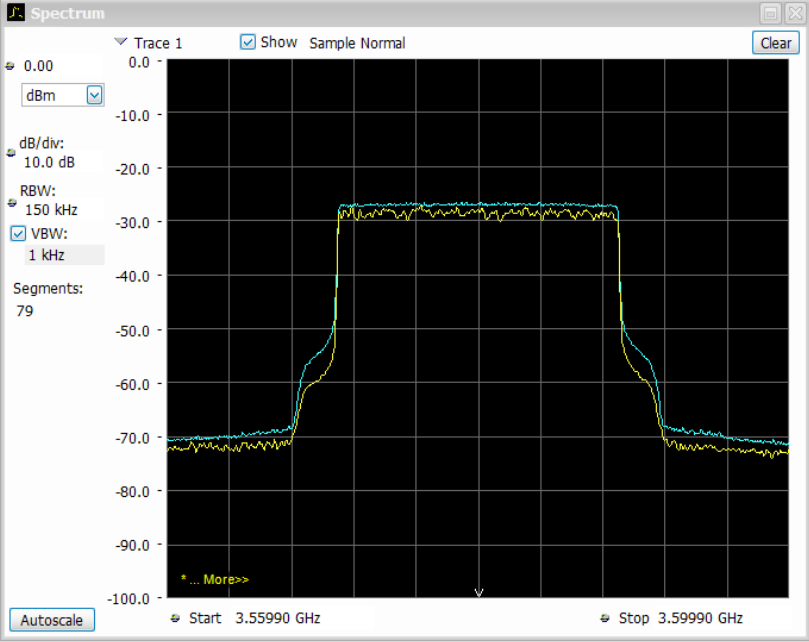}
\end{center}
\vspace{-2mm}
\caption
{Transmit signal spectrum at the BS antenna port, measured with spectrum analyzer through a splitter and attenuator.
}
\label{fig:spectrum}
\vspace{-5mm}
\end{figure} 

\noindent\textbf{RF Front End:}
\textcolor{black}{The output power of the USRPs alone is insufficient 
\textcolor{black}{for establishing reliable communications links over} more than a few 10s of meters; hence,  a PA in the transmit chain is necessary. Given our goal to communicate over 1 km, our calculations have shown that a high linearity PA with a gain of 30-45 dB is needed. The 15 W PA from Table I requires a heavy heatsink and dedicated power supply and is recommended for the fixed node. The 
UAV imposes constraints on size, weight, and power. The smaller and lighter 1 W PA can be mounted directly to the payload chassis for cooling. 
Based on the maximum output power of the USRP and any signal distortion, appropriate input power backoff 
to provide a clean transmission. 
}

\textcolor{black}{Our experimental license prohibits out-of-band transmissions. A well-specified low-pass filter in the transmit chain provides simplicity and frequency agility while 
mitigating potential 
RF harmonics. 
A wide bandpass filter (preselector) 
in the receive chain helps with minimizing interference while 
allowing frequency agility. The specific filters for the band of interest are provided in Table I and were selected because of their specifications and availability.}

\textcolor{black}{In the receive chain, a low noise amplifier (LNA) was anticipated, and experimentally verified, to 
improve the communications range. The power supply voltage for the LNA and 1 W PA are identical, thus saving weight on the UAV.}

\textcolor{black}{The USRP has internal variable gain amplifiers in the transmit and receive paths. They allow for a range of linear amplification \cite{Keith_ACM20}, but these gains need to be carefully adjusted to optimize the connection depending on the PA and LNA 
\textcolor{black}{specifications
}, 
the Tx-Rx antenna isolation, and the desired 
range. \textcolor{black}{In general, the USRP Tx gain is set to the highest value to provide maximum signal strength while accounting for sufficient backoff to not damage the PA and avoid 
signal distortion
, as shown in Fig.~\ref{fig:spectrum}. The USRP Rx gain is set 
to be able to receive weak signals while avoiding receiver saturation from the desired signal, its own Tx signal, and any other signals falling into the passband of the preselector.}}

\textcolor{black}{Because a duplexer would limit frequency agility, 
we choose to minimize self-interference 
by physically isolating the antennas both on the UAV and at the BS (Fig.~\ref{fig:AERPAWFIELD}). 
In order to ensure good aerial coverage and simplify deployment, we chose 
wideband omnidirectional antennas. }

\noindent\textbf{Processing Unit:}
\textcolor{black}{Selecting the processing unit and SDR should be done jointly and as a function of the SDR software. Because of the need for a small form factor, lightweight, and low power consuming processing unit for the UAV, we had to compromise the 10 Gbit network interface card (NIC). This left us with USB-driven USRPs and the powerful and compact Intel NUC10 computer (Table I), which has many USB 3.0 ports for driving multiple such USRPs. The NUC10 also has a modern WiFi chip, leveraged for experiment control.
For the SDR BS, 
a multicore 
processor with highest per-core clock rates, a WiFi chip, and a 10 Gbit NIC are essential to support the necessary network interfaces of Fig.~\ref{fig:Figure3}. 
A high-end Intel i7 or i9 processor offers a high clock rate per core, can be packaged in a small form factor, and has a lower power consumption 
compared to high-end processors used for servers or workstations, easing field deployment. }


\noindent\textcolor{black}{\textbf{Power:}
The choice of the power supply source 
for the mobile radio node 
will impact the UAV flight duration. Although it is possible to power the payload from the UAV power system, an additional battery will limit flight time loss in spite of the increased weight \cite{Tethered_Relay}. This payload battery is sized by calculating the maximum power requirements of the system from the component data sheets and multiplying by the desired flight time of 0.5 h. Recognizing that the maximum power case is a worst-case scenario, no additional safety margin is incorporated. This calculation yields a 3000 mAh battery as optimum. 
Payload and UAV batteries are swapped and recharged at the same time. A 2000 W
gasoline generator using 0.5 gallon of fuel can supply the ground node equipment plus displays shown in Fig. \ref{fig:AERPAWFIELD} for about 7 h.}

\subsection{Radio Network Software}


Both OAI and srsRAN have a fully functional 4G LTE protocol stack and are developing 5G systems in software. They support multiple SDRs and are popular among researchers and developers.
OAI is built using scripts for creating the appropriate executables, whereas srsRAN uses a standard process for building from source. 
\textcolor{black}{Both libraries run on Ubuntu, are well maintained, and new versions are regularly released.} 


\textcolor{black}{Each software has advantages and disadvantages that make them useful for different experiments. 
Different modules of each software may be used to take advantage of their features. 
But 
this poses compatibility and interoperability challenges that 
may include 
incompatible interfaces or parameters. 
We analyzed the options and found that srsEPC is compatible with either the OAI BS (eNB) or srsENB and either combination supports the srsUE. Open5GS is compatible with srsENB 
and srsUE and 
supports handover.
}

\textcolor{black}{When using different eNBs, the key parameters 
are those relating to the EPC connection, such as the mobility management entity 
IP address, 
tracking area, mobile country and network codes. Those parameters must match the EPC in order for proper communication to occur between them. 
Using an EPC from a different project, e.g. Open5GS with srsENB, requires that the UE information is entered into its database. 
RAN settings include LTE band or frequency and USRP gain. 
}


\subsection{Network Interfaces}
\noindent{\textbf{Experimental Network:}}
The experimental 
cellular network 
uses the wireless interface and networked computers for implementing the RAN and CN. The CN can be integrated with one BS but other BSs need to connect to it through a wired or wireless link. 
\textcolor{black}{When deploying multiple BSs, the backhaul is ideally implemented via fiber. 
For flexibility, simplicity, and to not have to rely on costly infrastructure, we recommend setting up a dedicated WiFi network with directional antennas for higher range. 
A potential future solution is the integrated access and backhaul, which is an emerging 
3GPP technology that future ground and ABSs will likely leverage~\cite{tafintsev2020aerial}}.

\noindent\textbf{Experiment Control: }
For the experiment control interface, a cellular control link is a viable option for most places, but requires a data plan and lightweight yet quality LTE USB dongles. This is AERPAW's choice. Below we 
present an alternative that works well 
with minimum dependencies and no additional payload. 

\begin{figure}[t]
\begin{center}
\end{center}
\includegraphics[width=0.49\textwidth]{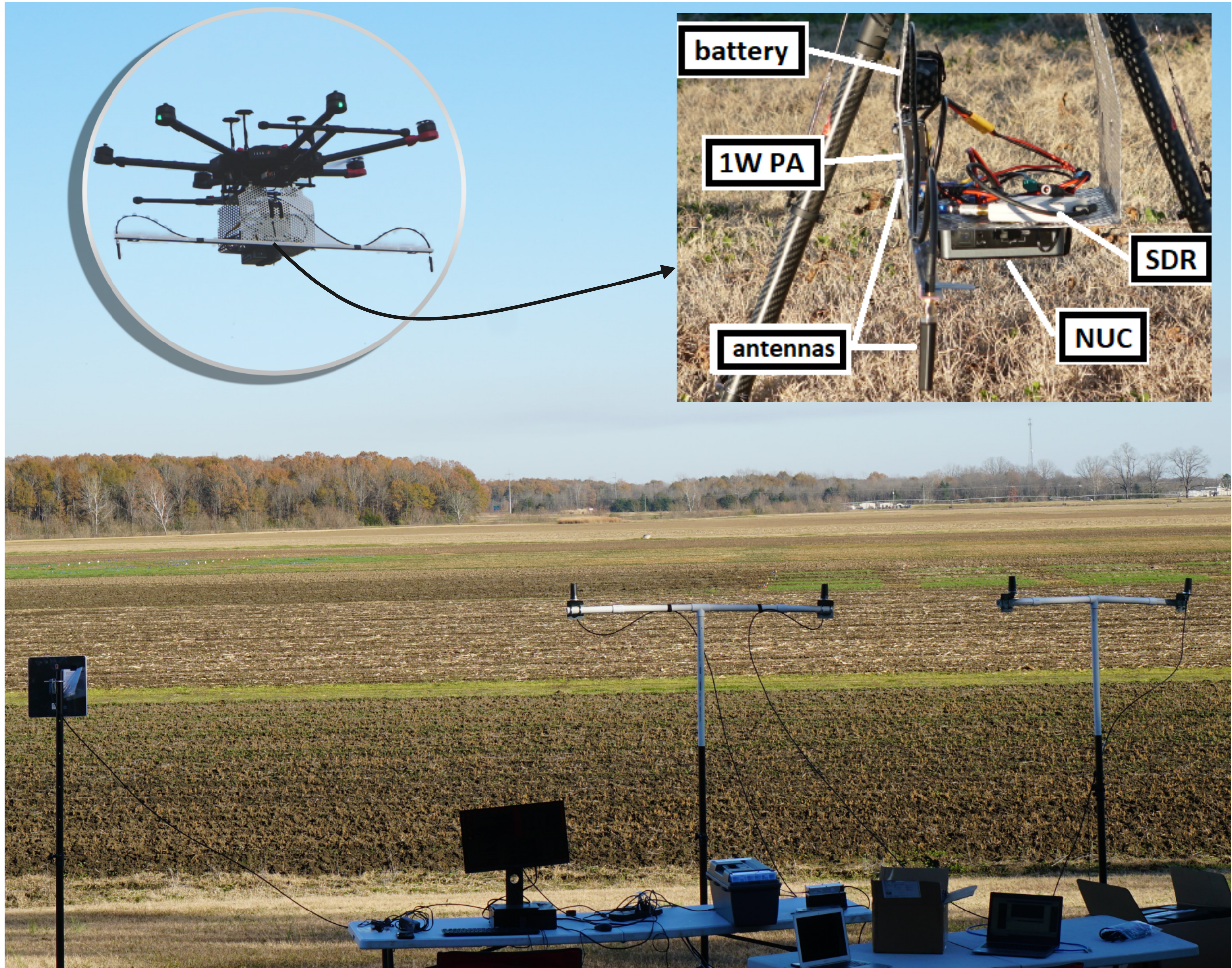}
\vspace{-4mm}
\caption{\textcolor{black}{Experiment with UAV and radio payload, directional 5~GHz antenna (bottom left), which communicates with the WiFi6 AP 700~m away from it in the field over which the UAV flies, SDR BS (middle), spectrum monitoring (right).}
}
\label{fig:AERPAWFIELD}
\vspace{-2mm}
\end{figure}

WiFi6, or IEEE 802.11ax, 
is widely available. 
It can be configured to operate in the 2.4 or 5 GHz unlicensed band and implements multi-user (MU)-MIMO for improving range and capacity. 
The BSs and UEs need to be within the range of the WiFi access point (AP) for continuous experiment control. 
COTS directional antennas are widely available for these bands and one can be collocated with each ground node for interconnecting BSs 
and implementing the 
required interfaces. 
The NUC10 has an integrated WiFi6 chip and antennas, and it is therefore mounted upside down on the outside of the UAV payload cage (Fig.~\ref{fig:AERPAWFIELD}). 


\subsection{UAS}
The weight of the payload and desired length of flight are the dominant factors determining UAV selection. Our platform (including payload battery) weighs 2.2 kg and we desire approximately 30 minutes of airborne experiment time. These requirements drive the need for a medium lift vehicle like the DJI Matrice 600. For flexibility \textcolor{black}{and exploratory research}
, we recommend manual control of the UAV flight during the experiment as opposed to waypoint-based flight path navigation. 
The control link is established through the standard 
RC provided by the UAV manufacturer and is independent of the previously discussed interfaces.  
Fig.~\ref{fig:AERPAWFIELD} shows our aerial 
node.
\begin{figure}[t]
\begin{center}
\vspace{-3mm}
\includegraphics[width=92mm]{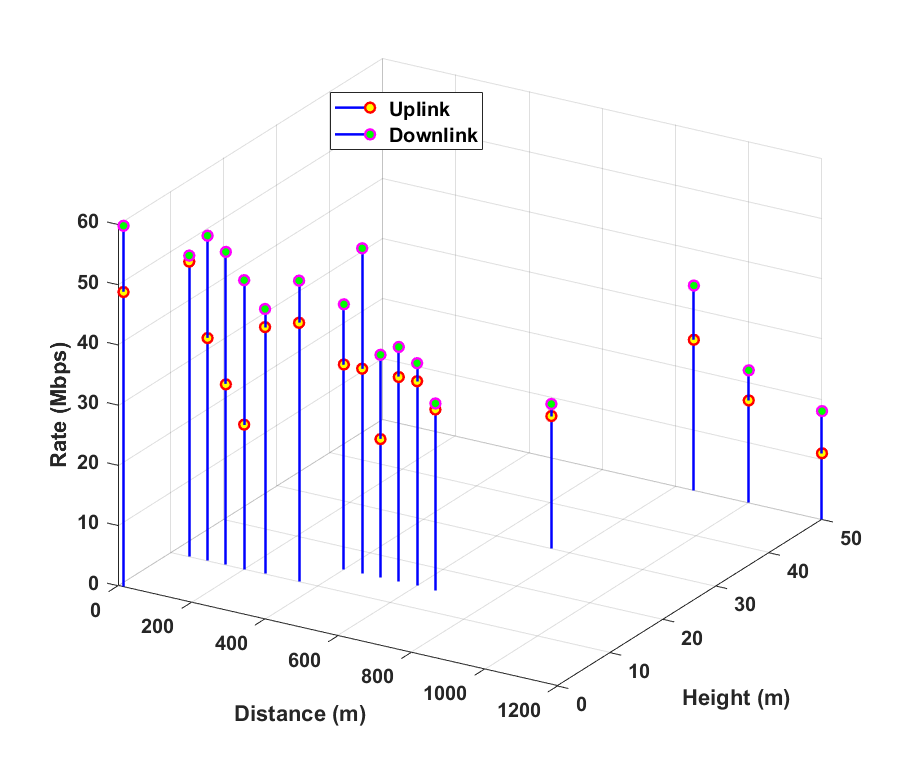}
\end{center}
\vspace{-5mm}
\caption{UL/DL throughput for 20 MHz FD-LTE using srsRAN at 3.5 GHz. The BS is at the origin and the UAV at (distance, height) with respect to it. 
}
\label{fig:Figure 20 MHz srsRAN}
\vspace{-5mm}
\end{figure} 

\subsection{Experimental Results}

Here we present initial  3D coverage measurements in a rural environment. 
We operate an LTE experiment in FDD mode (FD-LTE) in 3GPP Band 22 (C-Band), where we have a license and where 5G systems will operate. The fixed BS is 2.5 m above ground. The UAV flight is over an agricultural field with LOS to the BS (Fig. \ref{fig:AERPAWFIELD}).
We implement the eNB and UE with srsRAN and the EPC with open5GS. 
The USRP Tx and Rx gains are 72 and 47 for the eNB and 75 and 35 for the UE. 
\textcolor{black}{Fig. \ref{fig:spectrum} shows the spectrum of the eNB Tx signal, the LTE downlink (DL), at full DL resource allocation 
at the output of the 15 W PA for the above gains settings. After correcting for the splitter and attenuator losses 
in the spectrum analyzer path we obtain 
the average output power of 21 dBm at the eNB Tx antenna port. Equivalent measurements at the mobile node yield about 15 dBm average UE Tx power for the proposed 
configuration. The differences in Tx gains are due to the higher PA distortion at the eNB, whereas the differences in the Rx gains are due to the lower Tx-Rx antenna isolation at the UE.} 


Fig.~\ref{fig:Figure 20 MHz srsRAN} presents 
our uplink (UL) and DL throughput measurements. A UAV position close to the eNB at low altitude prove the successful attachment of the aerial UE (AUE) with the eNB and achieves nearly the highest theoretical rates for the UL and DL of 50 and 60 Mbps, respectively. 
About 25\% of the LTE frame is for control signaling which makes the theoretical maximum 50 and 75 Mbps, 
respectively. The throughput 
remains very high 
for more than 400 m horizontal distance from the BS at different UAV heights. 
The performance starts to degrade when the 
distance increases further as a result of 
the increased path loss, but 
is still well above 10 Mbps even beyond 1 km. 

\textcolor{black}{These results satisfy our goals to provide flexibility, performance, and enable scalable experiments. 
Compared to existing open-source SDR platforms and reported results, our platform offers high performance, range, and allows agility in frequency and flight trajectory. The platform can operate between 3.0 and 4.3~GHz, which can be further extended by changing the filters. All the deployed software is available for free download and use and is well maintained. All the hardware is widely available. The radio software and hardware choices are compliant with AERPAW and will be available through it for global researchers to use. One 
has still many choices, e.g. which UAV and frequency to use or how to implement the backhaul. 
This allows portability of our SDR platform with many customization and extension options for R\&D. 
} 


\subsection{Guidelines}
\textcolor{black}{
A few important guidelines 
for interested researchers are provided below. 
\begin{itemize}
    \item COTS SDR Tx power levels are low and there is often little isolation between the Tx and Rx signal paths. This requires signal, gain, and interference analysis, especially when using PAs. 
    \item The RF environment on the ground and in the air needs to be understood from the perspectives of the interference to and from the planned experiment. Adjacent channel signals passing the preselector of the SDR can cause aliasing or receiver saturation. Filters should therefore be used. 
    \item Processors, interfaces, the software environment, the SDR software processes, and the SDR and RF hardware need to be examined in concert and tested exhaustively to identify and eliminate bottlenecks. Baseline performance metrics and benchmarks need to be defined and RF and system testing processes established 
    to verify and calibrate the system configuration before each experiment.
\end{itemize}
}

\section{Challenges and Research Directions}
\label{sec:research}

The proposed platform has been developed to enable advanced research and help solving the challenges identified by industry and academia, including those described below.\\ 
\textbf{Interference:} The deployment of an 
aerial radio platform comes with various types of interference problems that can affect the network performance. 
In addition, the 
RF footprint of AUEs which typically have 
LOS channels 
with multiple 
ground BSs can 
severely degrade the capacity of the cellular network. 
Managing inter-cell interference (ICI) will be 
critical to enable coexistence between aerial and ground users. 
{The air-to-ground (A2G) ICI will exist between a UAV in one cell and cochannel interference between 
terrestrial and AUEs associated with nearby cells. }
The conventional ICI that has been studied for terrestrial communications will not be valid for solving the A2G ICI problem.  
Therefore, interference management 
is a critical research direction 
for UAV integration into cellular networks.\\ 
\textbf{Cell Association and Handover:} \textcolor{black}{\textcolor{black}{Because antennas have specific RF patterns and there is typically no signal blockage in the sky as there is on the ground that is leveraged to deploy cell towers and define cells boundaries for terrestrial users,} 
the traditional cell association and handover procedures will need to be revised to avoid frequent handovers and complex resource management. 
Aerial and terrestrial users may need to be treated differently. 
Advanced wireless technology, specifically beam-based access, can be leveraged for this.}\\ 
\textbf{Limited Airtime:} Battery-powered multi-rotor UAVs have an airtime of about thirty minutes. Different solutions can be conceived to address this limitation: (1) Design and deploy tethered UAVs where high endurance is needed and low mobility is possible. Research needs to make drone tethers more available, flexible, and safer to use \cite{Tethered_Relay}. (2) Coordinate UAVs to accomplish a mission, where one UAV can carry on the mission of another that needs a recharge. 
(3) Design and develop 
a combination of energy storage and charging technologies, as well as more energy-efficient UAV 
operation.\\ 
\textbf{UxNB:} UxNB is The 3GPP's term for UAVs taking on different roles in a cellular network. This includes the AUE, the ABS and the AR~\cite{3GPP_Stnds}. 
As we have flown the UE, we can fly the BS as well. In fact, the SDR software installed on all radio nodes 
implements the BS, CN, and UE. It is then a matter of executing the desired processes on the aerial or ground nodes. 
ABSs and ARs are popular concepts, but their gains and limitations need to be investigated.  
The coverage of ABSs will depend on the antennas used and the UAV height, among others. Research on 
deployment and radio resource management is much-needed for practical use cases in production-like environments. This is what the proposed SDR platform and the AERPAW testbed enable. 
    
    


\section{Conclusions}
\label{sec:conclusions}

\textcolor{black}{This paper has introduced the requirements and enabling technologies for 
implementing a modern  
cellular communications testbed for research on cellular networked UAVs. 
We provide detailed information about the SDR hardware and software, the RF front ends, the network, the UAS, and the power supply 
to successfully establish an open-source, software-defined 
broadband cellular network for R\&D. 
\textcolor{black}{Our measurement results show that the introduced platform can provide high throughput and the desired range between an AUE and a ground BS.}
These results offer early insights and will lead to \textcolor{black}{more research on} enhancing aerial 3D communications performance. The AERPAW testbed, deploying the SDRs presented here,  provides an at-scale research platform featuring multiple cells, 
micro and mm-wave radio transceivers, and multiple 
aerial and ground vehicles. 
It enables collecting 
data and 
evaluating different network configurations \textcolor{black}{for serving future UAS communications and networking needs}. \textcolor{black}{The open-source software-driven platform 
enables implementing 
the emerging open \textcolor{black}{RAN} architecture/interfaces and introducing network intelligence for 
research on
\textcolor{black}{6G} 
wireless.}} 
\balance

\section*{Acknowledgments}
This work 
was in part supported by 
NSF award CNS-1939334. We would like to thank Roshan KC, Joshua Moore, and Ajaya Dahal for their contributions to 
integration and testing.

\bibliographystyle{IEEEtran}
\bibliography{Refs}

\section*{Biographies}
\footnotesize
\noindent
\textbf{Aly Sabri Abdalla} (asa298@msstate.edu) is a PhD candidate in the Electrical and Computer Engineering (ECE) Department at Mississippi State University, Starkville, MS, USA. His research interests are on scheduling, congestion control and wireless security for vehicular ad-hoc and UAV networks.

\vspace{0.2cm}
\noindent
\textbf{Andrew Yingst} (aly112@msstate.edu) is pursuing a PhD degree in the ECE Department 
at Mississippi State University, Starkville, MS, USA. His research interests include long-range UAV communications and tethered UAV power systems.

\vspace{0.2cm}
\noindent
\textbf{Keith Powell} (kp1747@msstate.edu) is pursuing a PhD degree in the ECE Department 
at Mississippi State University, Starkville, MS, USA. His research interests include software radio platforms, UAV communications, and embedded systems.

\vspace{0.2cm}
\noindent
\textbf{Antoni Gelonch-Bosch} (antoni@tsc.upc.edu) is an associate professor in the Department of Signal Theory and Communications at 
Barcelona Tech, Castelldefels, Spain. His research interests include wireless communications, SDR systems, real-time processing, software abstraction, 
resource virtualization, and associated management.

\vspace{0.2cm}
\noindent
\textbf{Vuk Marojevic} (vuk.marojevic@msstate.edu) is an associate professor in ECE at Mississippi State University, Starkville, MS, USA. His research interests include mobile 
communications and wireless security with application to 
mission-critical networks and UAS. 

\end{document}